\documentclass[a4paper,11pt]{article}
\usepackage{pos}
\usepackage{aas_macros}

\newcommand{\er}[1]{\textcolor{black}{#1}}

\title{The Impact of Southern-Hemisphere Radio Blazar Observations on Neutrino Astronomy}

\author*[a,b]{F.~R{\"o}sch}
\author[b,a]{P.~Benke}
\author[a]{M.~Kadler}
\author[b]{E.~Ros}
\author[c]{R.~Ojha}
\author[d]{P.~G.~Edwards}
\author[a,b]{F.~Eppel}
\author[a,b]{J.~He{\ss}d{\"o}rfer}
\author[d]{J.~Stevens}

\affiliation[a]{Julius-Maximilians-Universit{\"a}t W{\"u}rzburg, Fakult{\"a}t f{\"u}r Physik und Astronomie, Institut f{\"u}r Theoretische Physik und Astrophysik, Lehrstuhl f{\"u}r Astronomie, Emil-Fischer-Str. 31, D-97074 W{\"u}rzburg, Germany}

\affiliation[b]{Max-Planck-Institut f{\"u}r Radioastronomie, Auf dem H{\"u}gel 69, \er{D-}53121 Bonn, Germany}

\affiliation[c]{NASA HQ, 300 E St SW, DC 20546-0002, Washington DC, USA}

\affiliation[d]{CSIRO Space and Astronomy, ATNF, PO Box 76, Epping NSW 1710, Australia}

\emailAdd{florian.roesch@physik.uni-wuerzburg.de}

\abstract{The origin of high-energy cosmic neutrinos detected by the IceCube observatory is a hotly debated topic in astroparticle physics. 
\er{There is growing evidence}
that some of these neutrinos can be associated with active galactic nuclei (AGN) and especially with blazars. Several recent studies have revealed a statistical correlation between radio-bright AGN samples and IceCube neutrino event catalogs. In addition, a growing number of individual high-energy neutrinos have been found to coincide with individual radio-flaring blazars. These observational results strongly call for high-quality, high angular-resolution radio observations of such neutrino-associated blazars to study their parsec-scale jet structures. TANAMI is the only large and long-term VLBI monitoring program 
\er{focused} on the Southern sky.
Within TANAMI, we put an emphasis on Southern IceCube neutrino candidate blazars at 2.3\,GHz and 8.4\,GHz. 
Here we present first results of the first high-quality, high angular-resolution VLBI observations of nine Southern-Hemisphere blazars that were associated to IceCube neutrino hotspots in the Southern sky. 
In the near future, the rapidly growing KM3NeT will complement IceCube 
\er{by being}
sensitive to high-energy neutrinos mainly from the Southern Hemisphere. This 
\er{will increase} 
the importance of Southern-Hemisphere radio monitoring programs of neutrino-associated blazars, like TANAMI.}

\ConferenceLogo{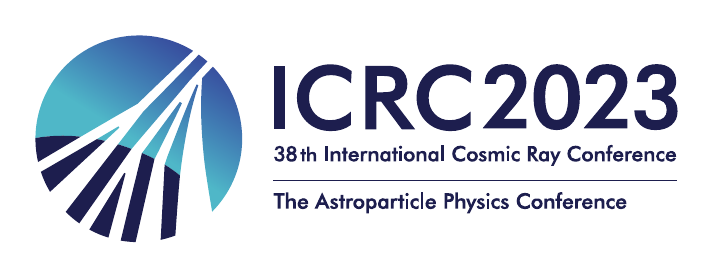}

\FullConference{%
38th International Cosmic Ray Conference (ICRC2023)\\
  26 July - 3 August, 2023\\
  Nagoya, Japan}

\begin{document}
\maketitle

\section{Introduction}
\label{sec:introduction}
\noindent
In recent years, there has been increasing evidence that some high-energy cosmic neutrinos detected by the IceCube observatory can be associated with active galactic nuclei (AGN), especially with blazars. The IceCube Collaboration \cite{IceCube2022} reported a correlation of neutrinos with known $\gamma$-ray emitters at a significance of $3.3\,\sigma$ with the largest contributions from the nearby active galaxy NGC\,1068 and the three blazars TXS\,0506+056, PKS\,1424+240 and GB6\,J1542+6129. 
\er{In addition,}
significant correlations have been claimed both between IceCube neutrinos and blazars from the Roma-5BZCat\footnote{\url{https://www.ssdc.asi.it/bzcat5/}} catalog \cite{Buson2022,Buson2023}, and with radio catalogs \cite{Hovatta2021,Plavin2020,Plavin2021,Plavin2023}. 
However, it has long been 
\er{shown}
that known $\gamma$-ray blazars can contribute only a limited fraction to the observed diffuse neutrino flux \cite{Aartsen2017}. 
A recent study \cite{IceCube2023a} found that this fraction might be as small as $<1\,\%$ and that the same is 
true for $\gamma$-undetected but radio-bright blazars from the radio fundamental catalog\footnote{\url{http://astrogeo.org/rfc/}} (RFC). 
However, there are still associations between high-energy cosmic neutrinos and individual flaring blazars. 
Indeed, in 2017 an association of the track-like muon neutrino event IC\,170922A with the $\gamma$-ray blazar TXS\,0506+056 
\er{was}
found 
\er{with}
a significance of $\sim3\sigma$ \cite{IceCube2018a}. 
Another example for such an association is the Southern-Hemisphere blazar PKS\,1424$-$418 which was found to be associated with the IceCube neutrino event IC\,35 (also known as 'BigBird') 
\er{with}
a significance of $\sim2\sigma$ \cite{Kadler2016}.

Blazars are radio-loud AGN hosting relativistic jets pointed close to our line of sight. Their emission is highly beamed and Doppler-boosted,
\er{making}
them variable broadband emitters from radio to $\gamma$-ray energies. Their spectral energy distribution (SED) shows a double-humped spectrum. While the first component corresponds to synchrotron radiation, there are two different types of models that are used to explain the high-energy emission of blazars: inverse Compton scattering and hadronic emission models. 
In the latter, the high-energy emission is produced by 
\er{the}
interactions of relativistic protons in the jet with soft ambient seed photons \cite{Mannheim}, and neutrino emission is naturally expected. The origin and properties of the seed photon fields are major open questions. 
On the one hand, the seed photon fields 
\er{may originate}
from external regions of the AGN. 
On the other hand, an attractive model \cite{Tavecchio2014,Tavecchio2015} 
\er{suggests}
that the high-energy protons of the fast inner spine of the blazar jet interact with soft target photons of the sheath, a slower jet layer surrounding the spine. Such spine-sheath structures can be revealed by radio observations using the technique of very-long-baseline interferometry (VLBI). 
Indeed, previous VLBI observations of the neutrino-associated blazars PKS\,1424-418, TXS\,0506+056 and PKS\,1502+106 have found indications of a limb-brightend structure, as predicted 
\er{by}
such spine-sheath models \cite{Kadler2016,Ros2020,Karamanavis2016}.

In this study we analyze 2.3\,GHz VLBI observations of a sample of nine Southern-Hemisphere blazars that are associated with IceCube neutrino hotspots, according to \cite{Buson2022}, to search for properties that might be characteristic for neutrino-emitting AGN. These observations represent the first high-quality, high angular-resolution VLBI observations of these nine blazars at 2.3\,GHz.

\section{TANAMI}
\label{sec:tanami}
\noindent
TANAMI (Tracking Active galactic Nuclei with Austral Milliarcsecond Interferometry) is the only large and long-term VLBI and multi-wavelength \er{blazar}
monitoring program 
in the Southern Hemisphere. 
It originally started in 2007 as a VLBI program 
\er{observing}
extragalactic jets south of $-30^\circ$ declination \cite{Ojha2010} and was then complemented with multi-wavelength observations at IR, optical/UV, X-ray and $\gamma$-ray energies and an additional radio flux density monitoring program using the Australian Telescope Compact Array (ATCA)\cite{Kadler2015}. There was also a close collaboration with the ANTARES neutrino telescope which was located in the Mediterranean Sea and therefore
\er{reached}
its highest sensitivity at declinations south of $-30^\circ$, 
\er{corresponding}
to the TANAMI sky \cite{Kadler2015}. 
From 2007 to 2020, TANAMI VLBI observations were performed at 8.4\,GHz and 22.3\,GHz using an array consisting of the five Australian Long Baseline Array (LBA) antennas, Parkes, ATCA, Hobart, Mopra and Ceduna, and an antenna in Hartebeesthoek, South Africa, as well as the 34\,m and 70\,m telescopes of the NASA Deep Space Network in Tidbinbilla. This array was further complemented by two German International VLBI Service (IVS) antennas, GARS at O’Higgins, Antarctica, and TIGO which was first at Concepci\'on, Chile, and later was moved to La Plata, Argentina \cite{Kadler2015}. 
Since 2020 the TANAMI sky has been 
\er{extended}
to the 
\er{entire}
Southern sky,
\er{which means}
that also blazars between $0^\circ$ and $-30^\circ$ declination 
\er{are}
included in the TANAMI sample. Furthermore, to 
detect faint TeV-detected sources and neutrino candidate blazars, 2.3\,GHz observations, using an array consisting of the five LBA antennas, the Hartebeesthoek telescope, the Tidbinbilla stations and three IVS antennas, Katherine and Yarragadee in Australia and Warkworth in New Zealand, have been added to the monitoring program.

\section{The Sample}
\label{sec:sample}
\noindent
In 2022, a sample of ten Southern-Hemisphere blazars from the Roma-5BZCat catalog were associated with IceCube neutrino hotspots in the Southern sky \cite{Buson2022}.
This sample includes five quasars, 
three BL\,Lac objects, and two blazars of uncertain classification (BZU). 
One source, 1814$-$637 (5BZU\,J1819$-$6345), has already been observed at 8.4\,GHz with TANAMI in 2008 and its morphology shows a compact symmetric structure 
\cite{Ojha2010} (see also \cite{Tzioumis2002,Morganti2011}). 
The other nine sources are likely blazars 
which we observed in August 2022 for the first time at VLBI resolutions at 2.3\,GHz using the TANAMI array. These nine sources are listed in Table~\ref{tab:imageparameters} and their high-image-fidelity brightness distributions on parsec scales are presented in this study (see Fig.~\ref{fig:images}). 

\section{TANAMI Observations at 2.3\,GHz}
\label{sec:observations}
\begin{figure}
    \centering
    \includegraphics[width=0.32\columnwidth]{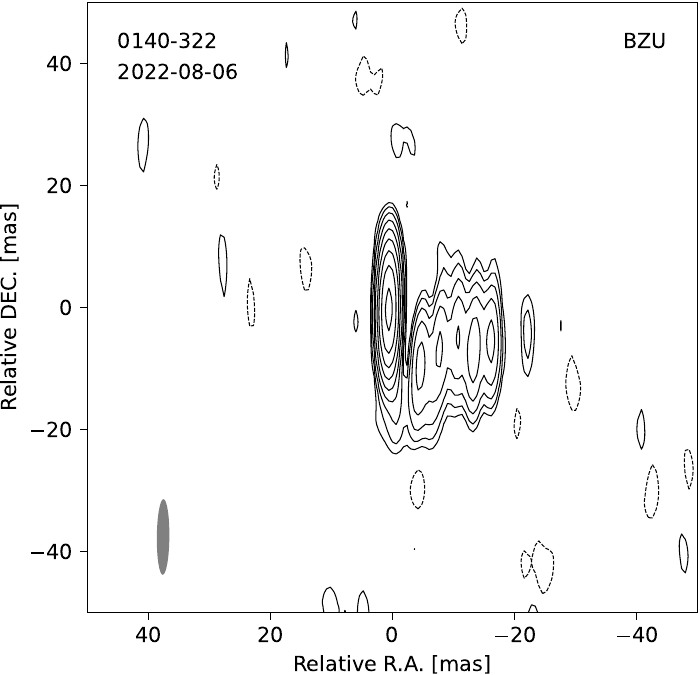}
    \includegraphics[width=0.32\columnwidth]{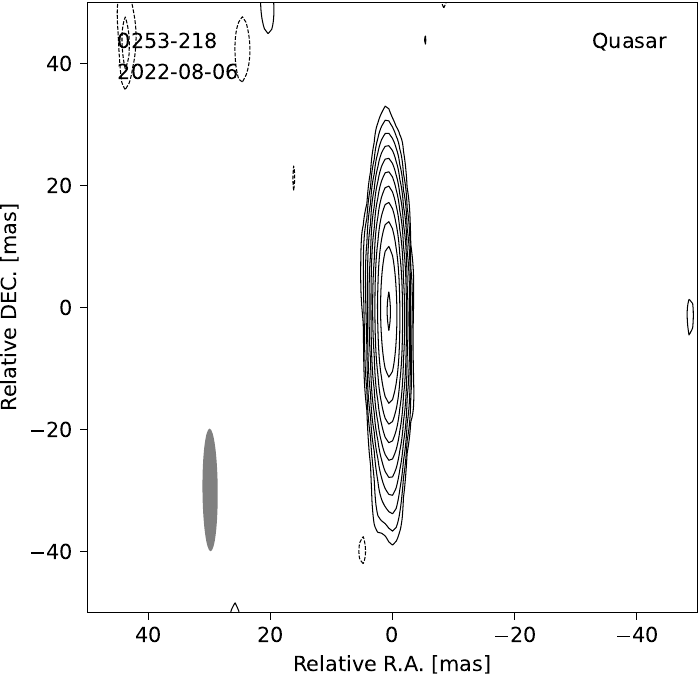}
    \includegraphics[width=0.32\columnwidth]{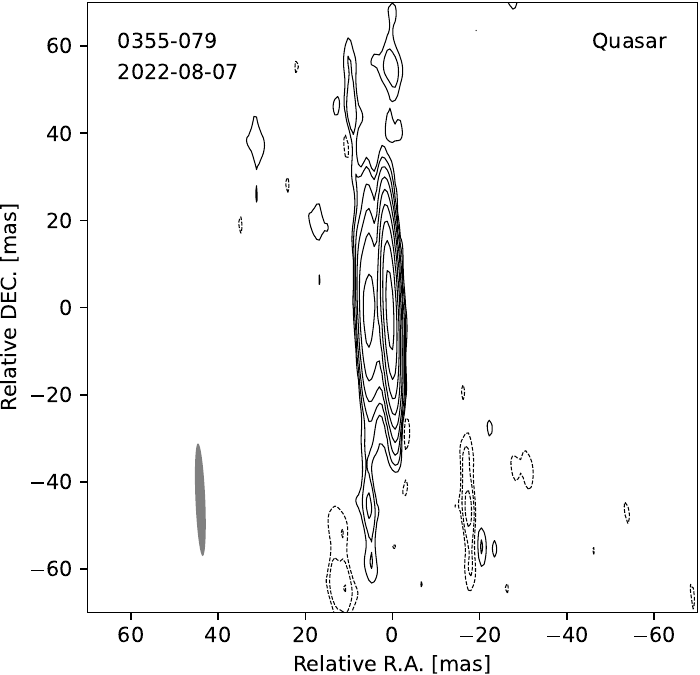}
    \includegraphics[width=0.32\columnwidth]{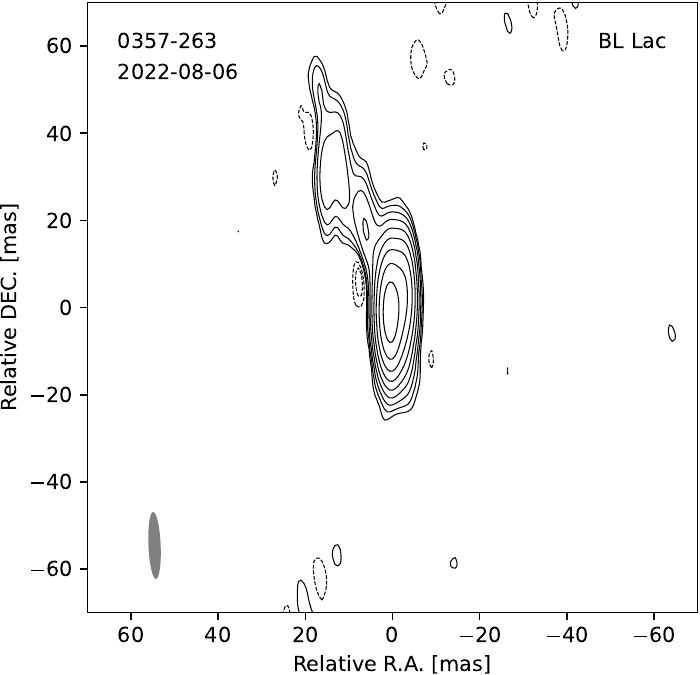}
    \includegraphics[width=0.32\columnwidth]{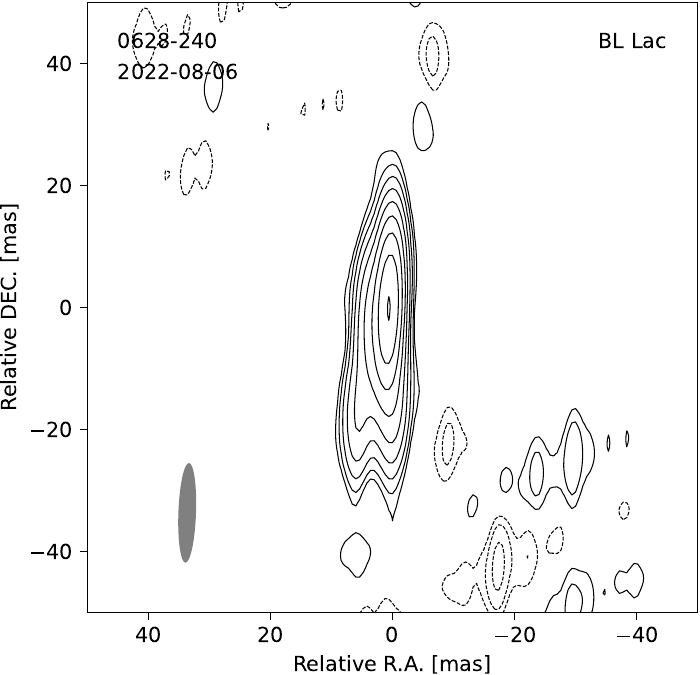}
    \includegraphics[width=0.32\columnwidth]{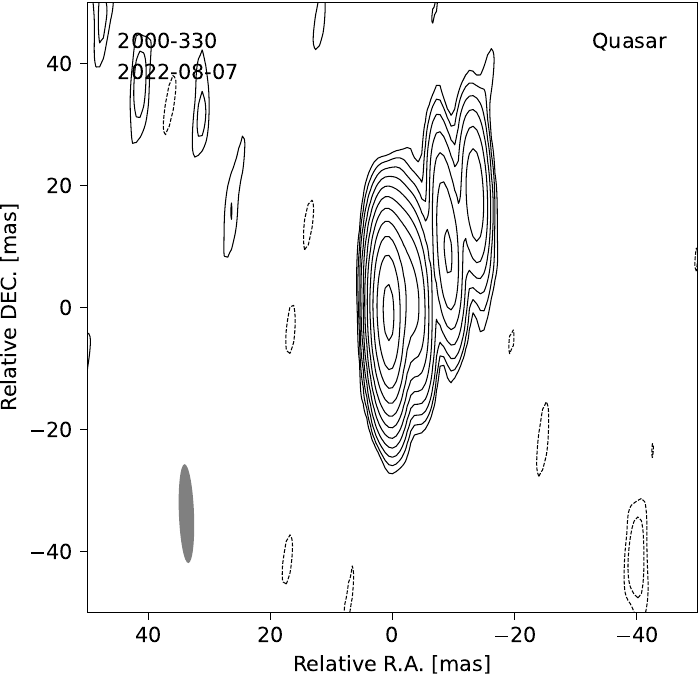}
    \includegraphics[width=0.32\columnwidth]{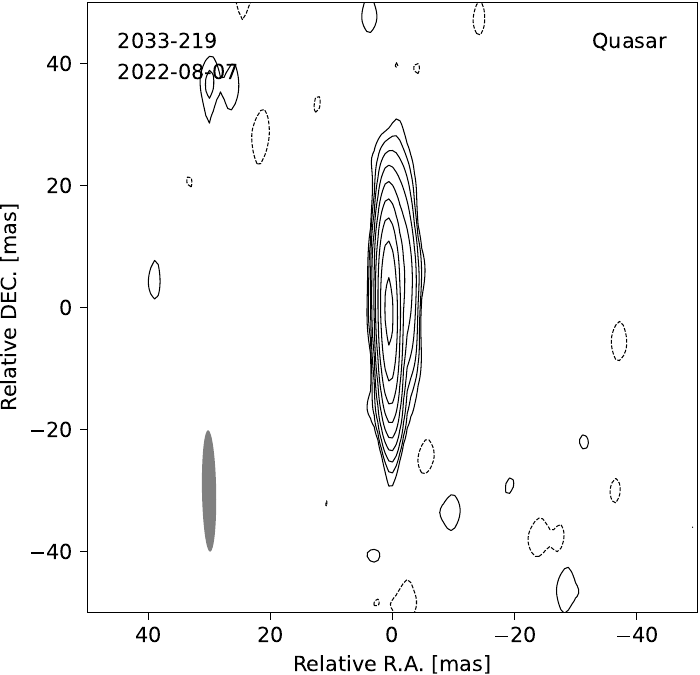}
    \includegraphics[width=0.32\columnwidth]{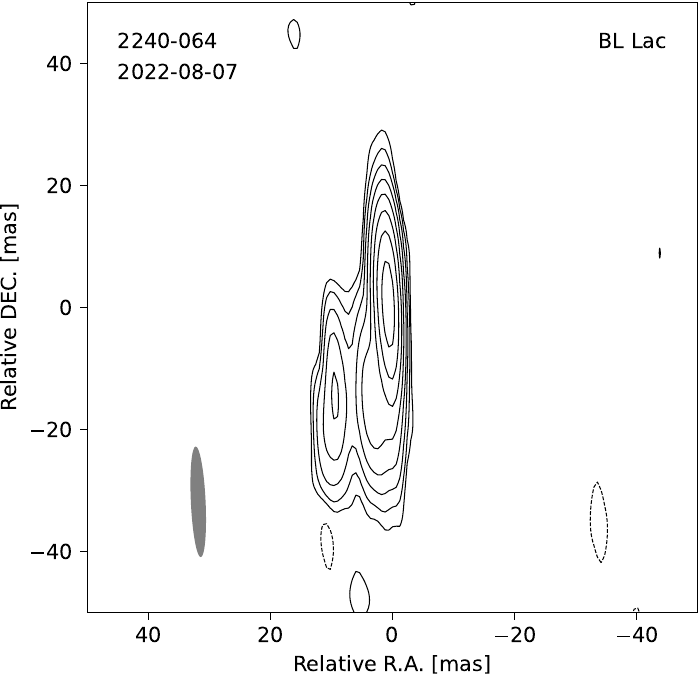}
    \includegraphics[width=0.32\columnwidth]{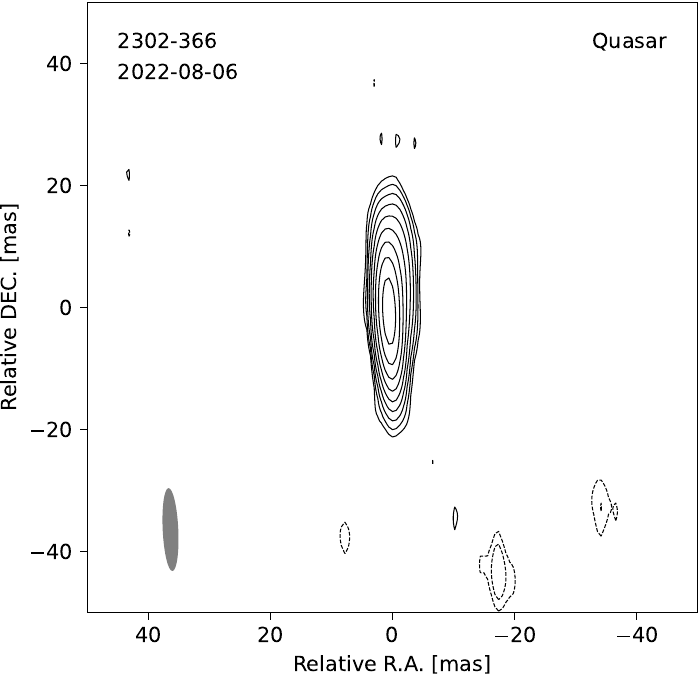}
    \caption{\small \sl Preliminary naturally weighted 2.3\,GHz TANAMI images of Southern-Hemisphere blazars 
    associated with IceCube neutrino hotspots according to \cite{Buson2022}. The contours start at $3\sigma$ and increase logarithmically by factors of 2. The grey ellipse in the bottom left corner of each image corresponds to the size of the beam. The beam parameters are listed in \er{Table}~\ref{tab:imageparameters}.}
    \label{fig:images}
\end{figure}

\noindent
To study the jet morphology and brightness temperature distribution of the nine sample sources presented in Sect.~\ref{sec:sample}, we observed them with the TANAMI array at 2.3\,GHz on August 6 and 7, 2022. The data were calibrated and imaged using the programs \texttt{AIPS} \cite{Greisen2003} and \texttt{DIFMAP} \cite{Shepherd2}, respectively, in a similar way as presented in \cite{Ojha2010}. The resulting naturally weighted images are shown in Fig.~\ref{fig:images} and their image parameters are listed in Table~\ref{tab:imageparameters}. The relative uncertainties of the flux densities are estimated to be 20\,\% \cite{Ojha2010}. Most of the sources are relatively faint with total integrated flux densities of around $100-200\,\mathrm{mJy}$ or below and are very compact showing a bright core component and faint jets. The quasar 2000$-$330 is the brightest source  with an integrated flux density of $\sim 600\,\mathrm{mJy}$ and a prominent jet to the northwest. 

\begin{table*}
\centering
\small
\begin{tabular}{@{}ccccccccc@{}}
\hline\hline
Name & $z$ & $S_\mathrm{tot}$ & $S_\mathrm{peak}$ & $\sigma_\mathrm{rms}$ & $b_\mathrm{maj}$ & $b_\mathrm{min}$ &   P.A. & $T_\mathrm{B}$ \\ 
B1950 &  &            [Jy] &         [Jy/beam] &            [mJy/beam] &            [mas] &            [mas] &  [deg] & [K] \\
(1) &   (2) &              (3) &               (4) &                   (5) &              (6) &              (7) &    (8) & (9) \\
\hline
0140$-$322 & 0.375 &            0.101 &             0.074 &                 0.077 &           12.4 &            2.06 & -0.552 & $>1.30\cdot10^{10}$ \\
0253$-$218 & 1.47  &            0.144 &             0.138 &                 0.042 &           20.1 &            2.50 & 0.457 & $9.01\cdot10^{10}$ \\
0355$-$079 & 1.05 &            0.176 &             0.142 &                 0.129 &           25.9 &            2.25 & 2.42 & $>4.63\cdot10^{11}$ \\
0357$-$263 & 1.47$^a$ &            0.277 &             0.138 &                 0.111 &           15.4 &            2.80 & 2.41 & $8.57\cdot10^{9}$ \\
0628$-$240 & >1.239$^b$ &            0.057 &             0.041 &                 0.051 &           16.3 &            2.89 & -2.04 & $>4.66\cdot10^{9}$ \\
2000$-$330 & 3.78$^c$ &            0.594 &             0.365 &                 0.096 &           16.2 &            2.48 & 2.21 & $1.44\cdot10^{12}$ \\
2033$-$219 & 2.299 &            0.170 &             0.155 &                 0.163 &           19.9 &            2.33 & 0.971 & $1.57\cdot10^{11}$ \\
2240$-$064 & 0.30 &            0.098 &             0.063 &                 0.112 &           18.2 &            2.42 & 2.66 & $3.67\cdot10^{9}$ \\
2302$-$366 & 0.962 &            0.034 &             0.030 &                 0.025 &           13.6 &            2.56 & 2.70 & $9.24\cdot10^{9}$ \\
\hline
\end{tabular}
\caption{\label{tab:imageparameters} \small \sl Image parameters and brightness temperatures of the observed neutrino candidate blazars. Col.(1): B1950 name; Col.(2): Redshift taken from \cite{Buson2022}, $^a$redshift taken from \cite{Drinkwater1997},  $^b$redshift taken from \cite{Shaw2013}, $^c$redshift taken from \cite{Peterson1982}; Col.(3): Integrated total flux density; Col.(4): Highest flux density per beam; Col.(5): Noise level per beam; Col.(6): FWHM of the major axis of the beam; Col.(7): FWHM of the minor axis of the beam; Col.(8): Position angle of the major axis of the beam with respect to the beam's centroid (measured north through east); Col.(9): Brightness temperature of the core component.}
\end{table*}

\begin{figure}
    \centering
    \includegraphics[width=0.7\columnwidth]{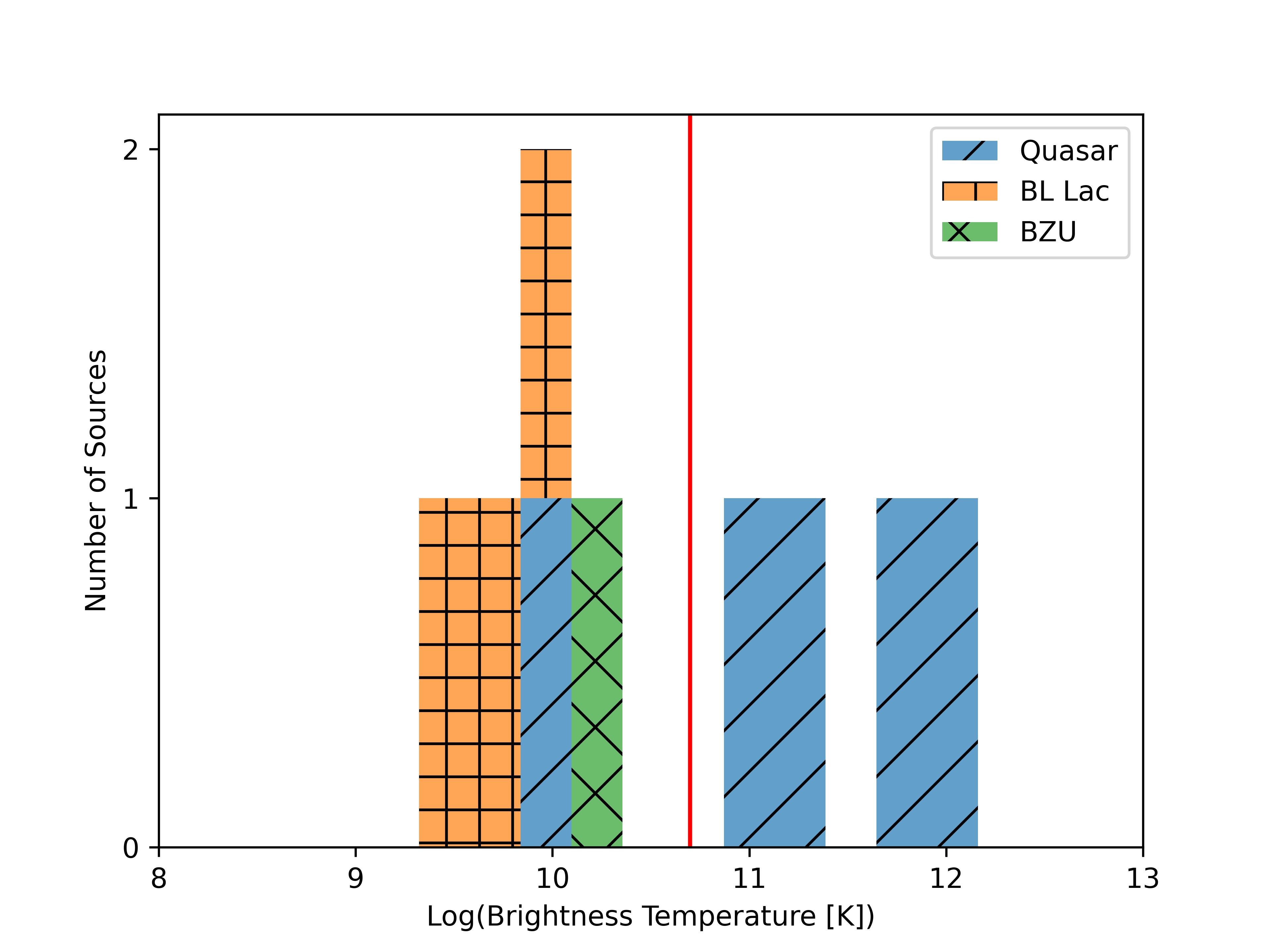}
    \caption{\small \sl Distribution of the brightness temperatures of the core components of the observed neutrino-associated blazars. The red line indicates the equipartition value of\, $T_\mathrm{eq}\approx 5\cdot10^{10}\,\mathrm{K}$ \cite{Readhead}.}
    \label{fig:tb}
\end{figure}

To study the distribution 
of the sources' core brightness temperatures, we fitted 2D Gaussian components to the fully imaged and self-calibrated visibilities using the \texttt{modelfit} command in \texttt{DIFMAP}. Based on the parameters of these Gaussian components we calculated the brightness temperatures of the core in the source's rest frame using Eq.~3 of \cite{Kovalev}. We also calculated the resolution limit of the core components based on Eq.~2 of \cite{Kovalev}. Whenever an axis of the fitted Gaussian components was smaller than the corresponding resolution limit, we considered 
\er{that}
component to be unresolved and used the corresponding resolution limit as an upper limit of its size to calculate a lower limit of the brightness temperature. However, the brightness temperatures of resolved core components should also be treated as lower limits, 
\er{since}
even smaller structures inside the VLBI cores could dominate 
\er{their} emission \cite{Ojha2010}. 
The derived brightness temperatures of the core components are listed in Table~\ref{tab:imageparameters} and their distribution is shown in Fig.~\ref{fig:tb}. 
\er{It can be seen} that the brightness temperatures of the three BL Lac objects in the sample are all below the equipartition value of $T_\mathrm{eq}\approx5\cdot10^{10}\,\mathrm{K}$ \cite{Readhead}, while four out of five quasars show brightness temperatures above the equipartition value and only one of these four quasars exceeds the inverse Compton limit of $\approx10^{12}\,\mathrm{K}$ \cite{Kellermann1}. Interestingly, this brightness temperature distribution is similar to that found for TeV-detected blazars observed at 2.3\,GHz within the TANAMI program (Benke et al. in prep.). 


\section{Conclusion and Outlook}
\label{sec:conclusion}
\noindent
We 
\er{have for the first time} 
observed a sample of nine blazars 
associated with IceCube neutrino hotspots by \cite{Buson2022} at 2.3\,GHz 
in the TANAMI program 
at parsec-scale resolutions and calculated 
\er{the brightness temperatures of
their core components}. 
The resulting distribution of these brightness temperatures is similar to the brightness temperature distribution of TeV-bright blazars observed with the TANAMI array at 2.3\,GHz (Benke et al. in prep.). 
TeV-detected high-peaked BL\,Lac objects tend to show low brightness temperatures,
\er{indicating} low Doppler factors of the radio-emitting plasma. 
To explain this, and at the same time the high Doppler factors 
\er{inferred}
for the same sources from higher energy observations, models 
\er{suggesting}
separate emission regions with different Doppler factors \er{are discussed} \cite[e.g.,][]{Ghisellini2005}. 
The comparability of the neutrino candidate blazars in this sample with respect to TeV blazars opens the interesting possibility to interpret neutrino production models in a similar scenario \cite[e.g.,][]{Tavecchio2014,Tavecchio2015}.
However, we need to observe a larger 
\er{sample} of neutrino candidate blazars 
to 
\er{test this hypothesis}
in more detail. 
In the future, the rapidly growing KM3NeT neutrino telescope 
\er{in the Mediterranean} Sea could provide such a sample of Southern-Hemisphere neutrino-associated blazars, since it is 
\er{particularly} 
sensitive to neutrino-emitting sources 
\er{in} 
the Southern sky \cite{KM3Net2016}. 
This underlines the importance of Southern-Hemisphere radio monitoring programs
\er{such as}
TANAMI. 

\acknowledgments
\noindent
The Long Baseline Array is part of the Australia Telescope National Facility (\url{https://ror.org/05qajvd42}) which is funded by the Australian Government for operation as a National Facility managed by CSIRO. FE, JH, MK, and FR acknowledge support from the Deutsche Forschungsgemeinschaft (DFG, grants 447572188, 434448349, 465409577). PB was supported through a PhD grant from the International Max Planck Research School (IMPRS) for Astronomy and Astrophysics at the Universities of Bonn and Cologne.
We acknowledge the M2FINDERS project from the European Research Council (ERC) under the European Union’s Horizon 2020 research and innovation programme (grant agreement No 101018682).

\bibliographystyle{JHEP}
\bibliography{bibtex.bib}


%
%
%

\end{document}